\documentclass[conference, a4paper, 10pt]{IEEEtran}
\usepackage[british,UKenglish]{babel}
\usepackage[utf8]{inputenc}
\usepackage{csquotes}
\usepackage[T1]{fontenc}
\usepackage{textcomp}
\usepackage{microtype}
\usepackage{ragged2e}
\usepackage{mathtools}      
\usepackage{amssymb}
\usepackage{amsfonts}
\usepackage[cmintegrals]{newtxmath}
\usepackage{ntheorem}
\usepackage{algorithm}
\usepackage{algorithmic}
\usepackage{comment}
\usepackage{cite}
\usepackage{fancyref}
\usepackage{enumitem}  
\usepackage[
    acronym,
    toc,
    nopostdot,
    nogroupskip,
    nonumberlist,
    ]{glossaries}
\glsdisablehyper
\newacronym{cf}{CF}{cell-free}
\newacronym{mu}{MU}{multi-user}
\newacronym[ plural = MIMO, firstplural = multiple inputs multiple outputs (MIMO) ]{mimo}{MIMO}{multiple input multiple output}
\newacronym{rrm}{RRM}{radio resource management}
\newacronym{ra}{RA}{resource allocation}
\newacronym{rl}{RL}{reinforcement learning}
\newacronym[ plural = PFs, firstplural = Pareto frontiers (PFs) ]{pf}{PF}{Pareto frontier}
\newacronym{hv}{HV}{hypervolume indicator}
\newacronym{hvi}{HVI}{hypervolume improvement}
\newacronym{ehvi}{EHVI}{expected hypervolume improvement}
\newacronym{mo}{MO}{multi-objective}
\newacronym{mc}{MC}{Monte Carlo}

\newacronym[ plural = GPs, firstplural = Gaussian processes (GPs) ]{gp}{GP}{Gaussian process}
\newacronym[ plural = ANN, firstplural = artificial neural networks (ANN) ]{ann}{ANN}{artificial neural network}
\newacronym{rf}{RF}{radio frequency}
\newacronym{tdd}{TDD}{time division duplex}

\newacronym{ue}{UE}{user equipment}
\newacronym{dl}{DL}{downlink}
\newacronym{ul}{UL}{uplink}
\newacronym{se}{SE}{spectral efficiency}
\newacronym{minlp}{MINLP}{mixed-integer nonlinear programming}
\newacronym{socp}{SOCP}{second order cone programming}
\newacronym{sca}{SCA}{successive convex approximation}
\newacronym{mmfpc}{MMF-PC}{max-min fairness power control}
\newacronym{pc}{PC}{power control}

\newacronym[ plural = APs, firstplural = access points (APs) ]{ap}{AP}{access point}
\newacronym[ plural = SINRs, firstplural = signal-to-interference-plus-noise ratios (SINRs) ]{sinr}{SINR}{signal-to-interference-plus-noise ratio}

\newacronym{bo}{BO}{Bayesian optimisation}

\usepackage{array}           
\usepackage{booktabs}
\usepackage{longtable}
\usepackage{tabularx}  
\usepackage{threeparttablex}
\usepackage{multirow}
\usepackage{caption}       
\usepackage{float}
\usepackage{placeins}
\usepackage{subcaption}
\usepackage{rotating}
\usepackage{stfloats}
\usepackage{hyperref}            
\hypersetup{
    colorlinks=true,
    linkcolor=blue,
    citecolor=red,
    urlcolor=magenta,
}
\usepackage{url}
\usepackage{bookmark}
\usepackage{graphicx}                
\graphicspath{{./figures/}}
\DeclareGraphicsExtensions{.pdf,.png,.jpeg,.jpg,.eps,.svg}
\usepackage[dvipsnames]{xcolor}
\usepackage{tikz}          
\usepackage{pgfplots}
\usepackage{tikz-cd}
\usepackage{bloques}
\usepackage{datatool}          
\usepackage{textmerg}
\usepackage{gitinfo2}              

\newcommand{\etal}{et al. }    
\newcommand{\mx}[1]{\mathbf{#1}}

\begin{document}

% ––– Title ––––––––––––––––––––––––––––––
\title{
    Cell-Free Data Power Control Via \\ Scalable Multi-Objective Bayesian Optimisation
}

% ––– Authors incl. alt. ––––––––––––––––––––––––––––––
\author{
    Sergey~S.~Tambovskiy$^{\dag\ddag}$,
    Gábor~Fodor~$^{\dag\ddag}$,
    Hugo~Tullberg$^{\dag}$ \\
    \small $^\dag$Ericsson Research, Stockholm, Sweden. E-mail: \texttt{Sergey.Tambovskiy|Gabor.Fodor|Hugo.Tullberg@ericsson.com}\\
    \small $^\ddag$KTH Royal Institute of Technology, Stockholm, Sweden. E-mail: \texttt{sergeyta|gaborf@kth.se}
}

% ––– Place title ––––––––––––––––––––––––––––––
\maketitle

% ––– Abstract ––––––––––––––––––––––––––––––
\begin{abstract}
    \Glsentrylong{cf} \glsentrylong{mu} \glsentrylong{mimo} networks are a promising alternative to classical cellular architectures, since they have the potential to provide uniform service quality and high resource utilisation over the entire coverage area of the network.
    To realise this potential, previous works have developed \glsentrylong{rrm} mechanisms using various optimisation engines.
    In this work, we consider the problem of overall ergodic \glsentrylong{se} maximisation in the context of \glsentrylong{ul}-\glsentrylong{dl} data \glsentrylong{pc} in \glsentrylong{cf} networks.
    To solve this problem in large networks, and to address convergence-time limitations, we apply scalable multi-objective \glsentrylong{bo}.
    Furthermore, we discuss how an intersection of multi-fidelity emulation and \glsentrylong{bo} can improve \glsentrylong{rrm} in \glsentrylong{cf} networks.
\end{abstract}
  
% ––– Keywords ––––––––––––––––––––––––––––––
\begin{IEEEkeywords}
    Cell-free networks, Bayesian optimisation, power allocation, power control, radio resource management.
\end{IEEEkeywords}

% ––– Sections ––––––––––––––––––––––––––––––
\section{Introduction}\label{intro}
\Gls{cf} multi-user \gls{mimo} networks are promising 6G architectures, since they have the potential to provide a surface uniform quality of service while achieving high capacity and maintaining high time, frequency and spatial resource utilisation \cite{
    ranjbarCellFreeMMIMOSupport2022}.  
Their advantages over traditional cellular and distributed \gls{mimo} networks include flexible deployment of \glspl{ap}, dynamic \gls{ue} clustering, small fronthaul load and improved coverage with consequential increase in throughput \cite{
    chenSurveyUsercentricCellfree2021,
    bragaJointPilotData2022,
    wuProportionalFairResourceAllocation2022}.
Naturally, such an ambitious setup gives rise to a multitude of challenging optimisation problems in the domain of \gls{rrm}.
These do not only include \gls{ra}, \gls{pc} and load balancing \cite{
    nasirDeepReinforcementLearning2021,
    bragaJointPilotData2022}, 
but also tasks that stem from closely supporting functionalities at the physical and medium access control layers.
Examples of such mechanisms include link adaptation \cite{
    saxenaReinforcementLearningEfficient2022}, 
controller design for joint \gls{ul}-\gls{dl} scheduling \cite{
    girgisPredictiveControlCommunication2021}
and pilot assignment \cite{
    vanchienJointPilotDesign2018,
    nguyenPilotAssignmentJoint2021}.

As demonstrated in the following works, some of these problems are still addressed by conventional methods.
Wu \etal \cite{
    wuProportionalFairResourceAllocation2022}
tackle the combined slot \gls{ra} and precoder design problem by decoupling them into \gls{ue} grouping and intra-group \gls{ra} sub-tasks.
The former is approached as a graph-based optimisation, while the latter utilises the proportional fair \gls{ra} scheme.
Braga \etal \cite{bragaJointPilotData2022} 
study the interplay between a centralised \gls{ul} pilot and data \gls{pc} and its effects on the inherent trade-off between \gls{se} and fairness in \gls{cf} user-centric cases.
In their setting, the proposed geometric programming alone is insufficient due to the non-convexity of both problems and additional \gls{sinr} constraints imposed by an energy budget.
To remedy that, the authors derive a combination of successive convex approximation and geometric programming, which allows the former to be used with convergence guarantees.
The concept of inter-slice \gls{rrm} is briefly discussed by Wang et al. \cite{
    wangInterSliceRadioResource2021}, 
where the authors compare online convex optimisation with both model-based and data-driven methods.
While the proposed inter-slice \gls{rrm} scheme is appealing, it may suffer from a high computational complexity in scenarios that include a large number of \glspl{ap} or require high \gls{ra} frequency.  

The increasing popularity of data-driven modelling methodology is motivated by the abundance of available network and sensory data, and recent advances in statistical machine learning.
For example, \gls{rl} has already been applied for solving various \gls{rrm} tasks.
In \cite{
    saxenaReinforcementLearningEfficient2022}, 
Saxena \etal propose a tuning-free link adaptation mechanism using latent Thompson sampling supported by an offline link model with a Bayesian update scheme.
Nasir \etal \cite{
    nasirDeepReinforcementLearning2021}, 
use a multi-agent \gls{rl} framework to solve
the problem of simultaneous \gls{dl} subband selection and \gls{pc}.
Their solution stands out, as it features distributed execution and scales better in terms of the number of subbands in comparison with solutions that address these problems separately.
Finally, Li \etal \cite{
    liDeepReinforcementLearningBased2021}
apply deep \gls{rl} to balance fairness and spectrum efficiency by controlling beam selection in a set of \gls{ap} antenna arrays.

Concurrently, \gls{rrm} tasks may exhibit properties that hinder conventional \gls{rl} approaches. Examples of such tasks include objective functions that lack analytic expressions and have high evaluation costs.
To operate in such conditions, \gls{bo} may be used.
An early work to apply \gls{bo} in such a setting, is reported by Dreifuerst \etal \cite{
    dreifuerstOptimizingCoverageCapacity2021},
where the authors identify the set of Pareto optimal solutions between \gls{ul} capacity and coverage in the sectors of a multi-cell network by jointly tuning the downtilt and the transmit power.
While that method provides comparable results with \gls{rl}, \gls{bo} does not reduce the risk of creating coverage gaps with consequential performance degradation.
Later, Maggi \etal \cite{
    maggiBayesianOptimizationRadio2021}
successfully apply \gls{bo} to \gls{ul} open loop \gls{pc} in the form of a tutorial on both topics.
It can also be argued that the authors' formulation of \gls{bo} in a dynamic \gls{rrm} setting does not take into account pre-existing solutions addressing both cold-start and kernel selection issues.
Eller \etal \cite{
    ellerLocalizingBasestationsEndUser2022}
solve a reverse task, where \glspl{ue} infer the positions of \glspl{ap} from timing-advance measurements in a multipath propagation environment, using \gls{bo}.
Tekgul \etal \cite{
    tekgulSampleEfficientLearningCellular2021} 
optimise antenna downtilt from the perspectives of outage probability minimisation and sum log rate maximisation.
The authors embed \gls{bo} into differential evolution to solve, what is conventionally called, the problem of experimental design.
It is noteworthy, that this joint framework can effectively be replaced by a single \gls{bo} engine.
In fact, applying a single \gls{bo} engine can not only result in improved convergence guarantees, but it also increases the interpretability of the model.
To extend \gls{bo} into large dimensional settings, Maddox \etal \cite{
    maddoxOptimizingHighDimensionalPhysics2021} 
leverage tensor-based \gls{gp} surrogates for \gls{bo} and demonstrate the performance using various benchmarks, including coverage optimisation \cite{
    dreifuerstOptimizingCoverageCapacity2021}.

Continuing the line of research that lies in the intersection of \gls{rrm} and \gls{bo}, in this work we consider the problem of optimising the total ergodic \gls{se} in the context of \gls{ul}-\gls{dl} data \gls{pc} \cite{
    nguyenPilotAssignmentJoint2021} in \gls{cf} networks.
First, we argue that the aforementioned multi-objective problem can be approached from a \gls{bo} standpoint.
Next, we apply an advanced version of the \gls{bo} engine, providing uncertainty estimation and ensuring support of both multi-objective and high-dimensional aspects of the setting. 
We conclude this work with a discussion about the intersection of multi-fidelity emulation and \gls{bo}, and how they can improve \gls{rrm} in \gls{cf} \gls{mimo} networks.

\textit{Notations:}
The upper and lower-bold letters denote vectors and matrices,
$(\cdot)^{H}$ -- Hermitian transpose, $\mathbb{E}[\cdot]$ -- expectation of a random variable and $\mathcal{C N}(\cdot, \cdot)$ -- $\mathbb{C}$ Gaussian distribution.
The index $\alpha$ substitutes \gls{ul} and \gls{dl} indices.

\section{System Setting}\label{system}
\begin{figure}[hb]
    \centering
    \includegraphics[scale=0.69, keepaspectratio]{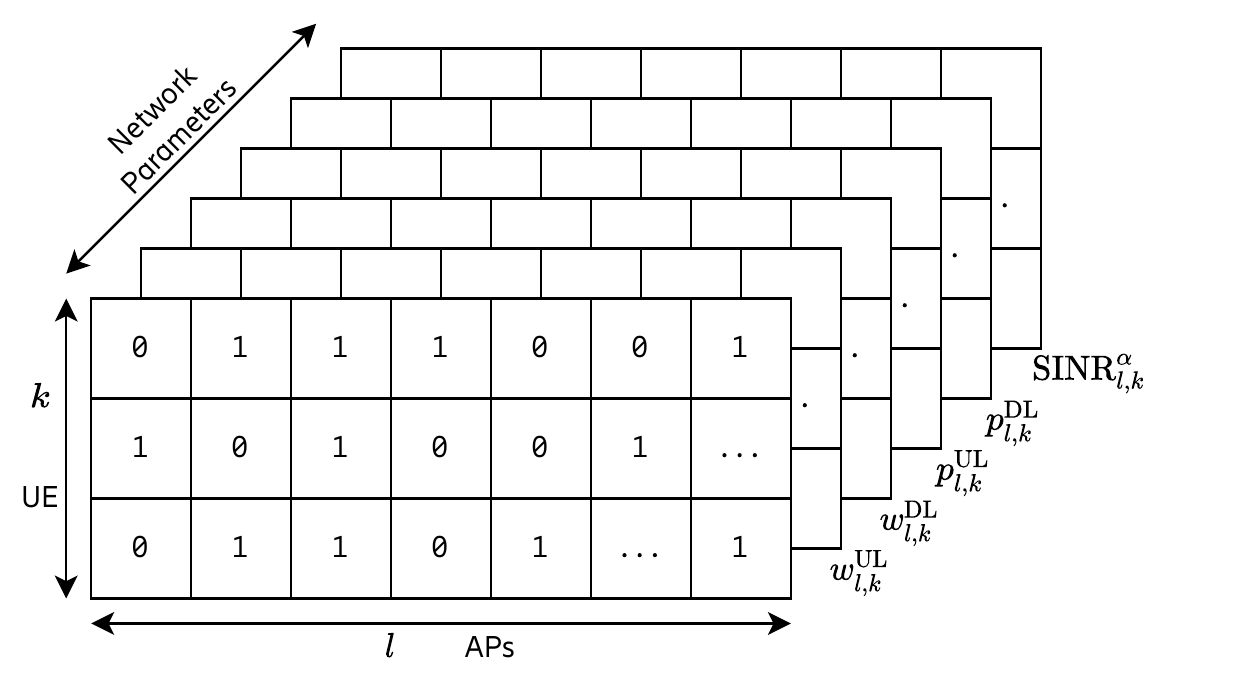}
    \caption{Tensor representation of a \gls{cf} network.}
    \label{fig-tensor}
\end{figure}

In this paper, we adapt the system model proposed
in \cite{nguyenPilotAssignmentJoint2021} for cellular \gls{mimo} networks to a \gls{cf} system setting.
Specifically, our system model consists of $L$ \gls{mimo} \glspl{ap}, each equipped with $M$ antennas, and $K$ single antenna \gls{ue}.
All units are placed randomly, in a uniform manner, on a limited Cartesian plane with a condition ensuring a minimum distance between any pair of units.
We also assume that each \gls{ue} can establish a \gls{tdd} link with each \gls{ap} and that 
latency constraints allow deployment of a centralised \gls{rrm} controller. 
We also introduce a concept of a tensor representation for the network.
Shown in Fig. \ref{fig-tensor}, the 3-dimensional array spans \gls{ue}, \glspl{ap} and parameters of \gls{tdd} links between aforementioned units.
Most notably, the first parameter depicts connectivity, where $1$ and $0$ indicate whether $k$-th \gls{ue} and $l$-th \gls{ap} are connected or not.
Consequentially it effects all parts of a system model containing sums over $k$ and $l$ indices.

Preserving the original definitions, in our system model $\tau_{c}$ denotes the length of coherence block, $\tau_{p}$ denotes the number of symbols for \gls{ul} training, thus $\tau_{c}-\tau_{p}$ is available for transmitting data symbols.
Fractions of the latter, $w_{l,k}^{\alpha}$, are used to tune the priorities of \gls{ul}-\gls{dl} transmissions between user $k$ and \gls{ap} $l$. 
The wireless channels $\mathbf{h}_{l,k} \in \mathbb{C}^{M}$ are assumed to have Rayleigh fading with $\mathbf{h}_{k,l} \sim \mathcal{C N}\left(\mathbf{0}, \mathbf{R}_{k,l}\right)$ distribution and correlation matrix $\mathbf{R}_{k,l} \in \mathbb{C}^{M \times M}$.
The channel information is assumed to be known or easy to estimate.
The \gls{tdd} links consist of \gls{ul} training and data transmission parts with definitions for
closed-form \gls{ul}-\gls{dl} ergodic \gls{se} (\ref{eq-erse}) and \gls{sinr} (\ref{eq-sinr}) defined in accordance with \cite{
    vanchienJointPilotDesign2018,
    nguyenPilotAssignmentJoint2021} 
and updated for \gls{cf} setting as follows:
\begin{equation}
    \label{eq-erse}
    R_{l, k}^{\alpha} = w_{l, k}^{\alpha}\left(1-\frac{\tau_{p}}{\tau_{c}}\right) \log _{2}\left(1+\operatorname{SINR}_{l, k}^{\alpha}\right)
\end{equation} 
\begin{equation}
    \label{eq-sinr}
    \begin{aligned}
        & \operatorname{SINR}_{l, k}^{\alpha} = A \mathbin{/} (I_C^{\alpha} + I_I^{\alpha} + \sigma_{\alpha}^{2}); \\
        & A = p_{l, k}^{\alpha}\left\|\boldsymbol{\psi}_{l, k}\right\|^{4} \operatorname{Tr}\left(\mathbf{R}_{l, k}^{} \mathbf{F}_{l, k}^{-1} \mathbf{R}_{l, k}^{}\right); \\
        & 
        I_C^{\text{UL}} = 
        \sum_{i, j \backslash l, k} p_{i, j}^{\text{UL}}\left|\boldsymbol{\psi}_{l, k}^{H} \boldsymbol{\psi}_{i, j}\right|^{2} \frac{\left|\operatorname{Tr}\left(\mathbf{R}_{i, j} \mathbf{F}_{l, k}^{-1} \mathbf{R}_{l, k}\right)\right|^{2}}{\operatorname{Tr}\left(\mathbf{R}_{l, k} \mathbf{F}_{l, k}^{-1} \mathbf{R}_{l, k}\right)}; \\
        & 
        I_C^{\text{DL}} = 
        \sum_{i, j \backslash l, k} p_{i, j}^{\text{DL}}\left|\boldsymbol{\psi}_{l, k}^{H} \boldsymbol{\psi}_{i, j}\right|^{2} \frac{\left|\operatorname{Tr}\left(\mathbf{R}_{i, j} \mathbf{F}_{i,j}^{-1} \mathbf{R}_{l, k}\right)\right|^{2}}{\operatorname{Tr}\left(\mathbf{R}_{i,j} \mathbf{F}_{i,j}^{-1} \mathbf{R}_{i,j}\right)}; \\
        & 
        I_I^{\text{UL}} = 
        \sum_{i, j \backslash l, k} p_{i, j}^{\text{UL}} \frac{\operatorname{Tr}\left(\mathbf{R}_{i, j} \mathbf{R}_{l, k} \mathbf{F}_{l, k}^{-1} \mathbf{R}_{l, k}\right)}{\operatorname{Tr}\left(\mathbf{R}_{l, k} \mathbf{F}_{l, k}^{-1} \mathbf{R}_{l, k}\right)}; \\
        & 
        I_I^{\text{DL}} = 
        \sum_{i, j \backslash l, k} p_{i, j}^{\text{DL}} \frac{\operatorname{Tr}\left(\mathbf{R}_{i, j} \mathbf{F}_{i, j}^{-1} \mathbf{R}_{i, j} \mathbf{R}_{l, k}\right)}{\operatorname{Tr}\left(\mathbf{R}_{i, j} \mathbf{F}_{i, j}^{-1} \mathbf{R}_{i, j}\right)}, \\
    \end{aligned}
\end{equation}
where $A$ denotes the array gain, $I_{C}^{\alpha}$ denotes the coherent interference (e.g. caused by pilot reuse), $I_{I}^{\alpha}$ is the incoherent interference and $\sigma_{\alpha}^{2}$ is the noise term.

The rest of the model, which contributes to (\ref{eq-erse}) follows the model proposed in \cite{nguyenPilotAssignmentJoint2021}, 
and is summarised as follows.
During \gls{ul} training, a number $K \leq \tau_{p} \leq K L$ of mutually orthogonal pilot signals $\boldsymbol{\psi}_{l, k}$ are reused among the \gls{ue} and contribute to \gls{ul} training pilot signal $\mathbf{Y}_{l, k} \in \mathbb{C}^{M \times \tau_{p}}$ at the \gls{ap} as in (\ref{eq-ulpiltr}).
\begin{equation}
    \label{eq-ulpiltr}
    \begin{aligned}
        & \mathbf{Y}_{l, k}=\sum_{i, j} \mathbf{h}_{i,j} \boldsymbol{\psi}_{i, j}^{H}+\mathbf{N}_{l, k}; \\
        & \hat{\mathbf{h}}_{l, k}=\left\|\boldsymbol{\psi}_{l, k}\right\|^{2} \mathbf{R}_{l, k}\left(\mathbf{F}_{l, k} \right)^{-1} \mathbf{Y}_{l,k} \boldsymbol{\psi}_{l,k}; \\
        & \hat{\mathbf{h}}_{l, k} \sim \mathcal{C N}\left(\mathbf{0},\left\|\boldsymbol{\psi}_{l, k}\right\|^{4} \mathbf{R}_{l, k} \mathbf{F}_{l, k}^{-1} \mathbf{R}_{l, k}\right); \\
        & \mathbf{F}_{l, k}=\sum_{i, j} \mathbf{R}_{i, t}\left|\boldsymbol{\psi}_{i, j}^{H} \boldsymbol{\psi}_{l, k}\right|^{2}+\sigma_{\text{UL}}^{2}\left\|\boldsymbol{\psi}_{l, k}\right\|^{2} \mathbf{I}_{M},
    \end{aligned}
\end{equation}
where $\mathbf{N}_{l, k} \in \mathcal{C N}\left(\mathbf{0}, \sigma_{\text{UL}}^{2}\mx{I}_{M \tau_p}\right)$ is the \gls{ul} noise matrix with its elements sampled from $\mathcal{C N}\left({0}, \sigma_{\text{UL}}^{2}\right)$ distribution and $\hat{\mathbf{h}}_{l, k}$ are channel estimates obtained by MMSE estimation \cite{
        vanchienJointPilotDesign2018, 
        nguyenPilotAssignmentJoint2021}.
The resulting channel estimates shape linear processing vectors for data transmission.
We make an assumption, that the timing difference between \gls{ue} \gls{ul} transmissions is insignificant, and they can be seen as practically simultaneous.  

During data reception by the \gls{ap}, the \gls{ul} signal $\mathbf{v}_{l, k}^{H} \mathbf{y}_{l}$ is detected by applying maximum-ratio combining vector $\mathbf{v}_{l, k}$ (\ref{eq-ul-data})
\begin{equation}
    \label{eq-ul-data}
    \begin{aligned}
        & \mathbf{v}_{l, k}^H \mathbf{y}_{l}=\sum_{i, t} \sqrt{p_{i, j}^{\text{UL}}} \hat{\mathbf{h}}_{l, k}^{H} \mathbf{h}_{i, j} s_{i, j}+\hat{\mathbf{h}}_{l, k}^{H} \mathbf{n}_{l}, \  \mathbf{v}_{l, k}=\hat{\mathbf{h}}_{l, k}; \\  
        & \mathbf{y}_{l}=\sum_{i, j} \sqrt{p_{i, j}^{\text{UL}}} \mathbf{h}_{i, j} s_{i, j}+\mathbf{n}_{l}, 
    \end{aligned}
\end{equation}
where $\mathbf{y}_{l}$ is a received signal with $s_{l, k}$ complex data, $p_{l,k}^{\mathrm{UL}}$ -- transmit data power and $\mathbf{n}_{l} \sim \mathcal{C N}\left(\mathbf{0}, \sigma_{\text{UL}}^{2} \mathbf{I}_{M}\right)$ -- noise term.
In turn, \gls{ue} receives a combined signal $r_{l, k}$ from all \glspl{ap} assigned to it, based on maximum ratio precoding vector $\mathbf{W}_{l, k}$.
\begin{equation}
    \label{eq-dl-data}
    \begin{aligned}
        & r_{l, k}=\sum_{i, j} \sqrt{p_{i, j}^{\text{DL}}}\left(\mathbf{h}_{l, k}^{i}\right)^{H} \mathbf{w}_{i,j} q_{i,j}+n_{l, k}; \\
        & \mathbf{w}_{l, k} = \hat{\mathbf{h}}_{l, k} \mathbin{/} {\sqrt{\left\|\boldsymbol{\psi}_{l, k}\right\|^{4} \operatorname{tr}\left(\mathbf{R}_{l, k}^{l} \mathbf{F}_{l, k}^{-1} \mathbf{R}_{l, k}^{l}\right)}}; \\
        & \mathbf{x}_{l}=\sum_{j=1}^{K} \sqrt{p_{l, j}^{\text{DL}}} \mathbf{w}_{l, j} q_{l, j},
    \end{aligned}
\end{equation}
where $\mathbf{x}_{l}$ is a transmitted signal with $q_{l, k}$ complex data, $p_{l, k}^{\mathrm{UL}}$ -- transmit data power and $n_{l, k} \sim \mathcal{C N}\left(0, \sigma_{\mathrm{dl}}^{2}\right)$ -- noise term.
Both (\ref{eq-ul-data}) and (\ref{eq-dl-data}) contribute directly to definition of \gls{se} in (\ref{eq-erse}).

\section{Constrained Data Power Allocation}\label{probfo}
In this part we introduce a problem of data \gls{pc} for the total max-min sum \gls{se} per user fairness under \gls{ul} and \gls{dl} transmit power constraints.
The fractions $w_{l,k}^{\alpha}$, introduced earlier in Section \ref{system}, from a system perspective serve not only as ratios between \gls{ul}-\gls{dl} symbol sequences, but also as prioritisations for each link established between $k$-th \gls{ue} and $l$-th \gls{ap}.
We address the problem (\ref{eq-problem-1}) in its original formulation \cite{nguyenPilotAssignmentJoint2021}
\begin{equation}
    \label{eq-problem-1}
    \begin{aligned}
        \max_{w_{l, k}^{\alpha}; \ p_{l, k}^{\alpha} \geq 0} \ \ \min_{(l, k)} \quad & R_{l, k}^{\text{UL}}+R_{l, k}^{\text{DL}} \\
        \textrm{s.t.} \quad & p_{l, k}^{\text{UL}} \leq P_{\max , l, k}^{\text{UL}}, \forall l, k \\
        & \sum_{k=1}^{K} p_{l, k}^{\text{DL}} \leq P_{\max , l}^{\text{DL}}, \forall l\text{,}    \\
    \end{aligned}
\end{equation}
where $P_{\max , l, k}^{\text{UL}}, P_{\max , l}^{\text{DL}}$ are maximum powers that each \gls{ue} and \gls{ap} can allocate to in the \gls{ul} and \gls{dl}, respectively.

Because this task falls into the combinatorial domain over a large scalable parameter space, previous works applied various heuristics in combination with conventional optimisation methods. These methods treat the \gls{ul}-\gls{dl} terms of the objective function separately or use sequential iterations between pilot assignment and \gls{pc}.
In this work, we relax the constraint on $\boldsymbol{\psi}_{l, k}$ effectively leaving only \gls{pc} for optimisation.
In the context of \gls{cf} \gls{mimo} networks, we must also ensure good scalability of the optimisation method as well as a good interpretability with uncertainty estimation for system simulations.
\Gls{bo} allows to address all of the aforementioned points.

\section{Multi-Objective Bayesian Optimisation}\label{gpmobo}
\subsection{Preliminaries}
\Gls{bo} is an efficient global black-box optimisation tool \cite{maggiBayesianOptimizationRadio2021}, that employs a probabilistic surrogate model in conjunction with an acquisition function to address the exploration-exploitation trade-off.
This methodology has been successfully developed over the years to support multi-objective cases and to a lesser extent problems with noisy observations and parallelisation requirements. 

Based on the formulation of constrained data \gls{pc} (\ref{eq-problem-1}), our objective, formally, is to find a set of optimal designs $\boldsymbol{x}$ over the bounded set $\mathcal{X}$ that maximise multiple objectives $\boldsymbol{f}(\boldsymbol{x})$.
Here, \gls{mo} optimisation aims to identify a set of Pareto-optimal objective trade-offs.
For example, a solution $\boldsymbol{f}(\boldsymbol{x})=\left[f^{(1)}(\boldsymbol{x}), \ldots, f^{(T)}(\boldsymbol{x})\right]$ dominates another solution $f(x) \succ f\left(x^{\prime}\right)$ if $f^{(t)}(\boldsymbol{x}) \geq f^{(t)}\left(\boldsymbol{x}^{\prime}\right)$ for $t=1, \ldots, T$ and $\exists t \in\{1, \ldots, T\}$ s.t. $f^{(t)}(\boldsymbol{x})>f^{(t)}\left(\boldsymbol{x}^{\prime}\right)$.
Then the \gls{pf} is defined as $\mathcal{P}^{*}=\left\{\boldsymbol{f}(\boldsymbol{x}): \boldsymbol{x} \in \mathcal{X}, \nexists \ \boldsymbol{x}^{\prime} \in \mathcal{X}\right.$ s.t. $\left.\boldsymbol{f}\left(\boldsymbol{x}^{\prime}\right) \succ \boldsymbol{f}(\boldsymbol{x})\right\}$ with a set of optimal designs as $\mathcal{X}^{*}=\left\{\boldsymbol{x}: \boldsymbol{f}(\boldsymbol{x}) \in \mathcal{P}^{*}\right\}$ .
The \Gls{pf} often consists of an infinite set of points and \gls{mo} optimisation algorithms aim to identify a finite approximate \gls{pf} $\mathcal{P}$.
The quality of \gls{pf} is the hypervolume of the region of objective space that is dominated by the \gls{pf} and bounded from below by a reference point \cite{daultonParallelBayesianOptimization2021}.

Within the aforementioned framework each objective must be modelled as an independent surrogate function.
\Glspl{gp} are typically used as such for \gls{bo} due to their well-calibrated predictive uncertainty.
Another part of \gls{bo} is an acquisition function $\alpha(\cdot)$, that specifies value of evaluating a set of new points based on the surrogate's predictive distribution.
Surrogates substitute original objectives $\boldsymbol{f}$ with fast and computationally cheap evaluations.
This allows to effectively run numerical optimisation methods to find $\boldsymbol{x}^{*}=\arg \max _{\boldsymbol{x} \in \mathcal{X}} \alpha(\boldsymbol{x})$ in the loop with a selection of new points for \gls{bo} to update the model.
Related research works contain many types of acquisition, which are chosen depending on the structure of the underlying problem \cite{wilsonMaximizingAcquisitionFunctions2018, gramacySurrogatesGaussianProcess2020}. 
Some of acquisition functions are probability of improvement, entropy search, upper confidence bound, and \gls{ehvi}, where the latter is used in this work.

\subsection{Expected Hypervolume Improvement}
We first recall that the explanation of parallel \gls{mo} \gls{bo} solutions, devised by Daulton \etal \cite{daultonParallelBayesianOptimization2021}, 
requires the definitions of the \gls{hv} and \gls{hvi}.

The \Gls{hv} of a finite approximate \gls{pf} $\mathcal{P}$ is the $T$-dimensional Lebesgue measure $\lambda_{T}$ of the space dominated by $\mathcal{P}$ and bounded from below by a reference point:

\begin{equation}
    \boldsymbol{r} \in \mathbb{R}^{T}: \mathrm{HV}(\mathcal{P} \mid \boldsymbol{r})=\lambda_{T}\left(\bigcup_{\boldsymbol{v} \in \mathcal{P}}[\boldsymbol{r}, \boldsymbol{v}]\right),    
\end{equation}

\noindent where $[\boldsymbol{r}, \boldsymbol{v}]$ denotes the hyper-rectangle bounded by vertices $\boldsymbol{r}$ and $\boldsymbol{v}$.
Then, the \Gls{hvi} of a {\em set of points} $\mathcal{P}^{\prime}$ with respect to an existing approximate \gls{pf} $\mathcal{P}$ and reference point $\boldsymbol{r}$ are defined as $\mathrm{HVI}\left(\mathcal{P}^{\prime} \mid \mathcal{P}, \boldsymbol{r}\right)=\mathrm{HV}\left(\mathcal{P} \cup \mathcal{P}^{\prime} \mid \boldsymbol{r}\right)-\mathrm{HV}(\mathcal{P} \mid \boldsymbol{r})$.

Since we approach our problem from black-box optimisation perspective, function values at unobserved points are unknown together with \gls{hvi} of an out-of-sample point. 
\Gls{bo} remedies this issue by \gls{gp} surrogates that provide a posterior distribution $p(\boldsymbol{f}(\boldsymbol{x}) \mid \mathcal{D})$ over function values for each $\boldsymbol{x}$.
This, in turn, allows to compute \gls{ehvi} acquisition function analytically (\ref{eq-ehvi-a}) or using \gls{mc} integration (\ref{eq-ehvi-b}):

\begin{equation}
    \label{eq-ehvi-a}
    \alpha_{\mathrm{EHVI}}(\boldsymbol{x} \mid \mathcal{P})=\mathbb{E}[\mathrm{HVI}(\boldsymbol{f}(\boldsymbol{x}) \mid \mathcal{P})]
\end{equation}

\begin{equation}
    \label{eq-ehvi-b}
    \begin{aligned}
        & \alpha_{q \mathrm{EHVI}}\left(\mathcal{X}_{\text {cand }} \mid \mathcal{P}\right) \approx \hat{\alpha}_{q \mathrm{EHVI}}\left(\mathcal{X}_{\text {cand }} \mid \mathcal{P}\right) = \\ 
        & ~~= \frac{1}{N} \sum_{i=1}^{N} \operatorname{HVI}\left(\tilde{\boldsymbol{f}}_{i}\left(\mathcal{X}_{\text {cand }}\right) \mid \mathcal{P}\right),
    \end{aligned}
\end{equation}
where $\tilde{\boldsymbol{f}}_{i} \sim p(\boldsymbol{f} \mid \mathcal{D})$ and $\mathcal{X}_{\text {cand }}=\left\{x_{j}\right\}_{i=j}^{q}$.

\subsection{Multi-Objective Noisy Expected Hypervolume Improvement}
Because conventional \gls{bo} does not scale well with number of objectives, and in such setting does not have converge guarantees, Daulton \etal \cite{daultonParallelBayesianOptimization2021} address that by deriving a Bayes-optimal \gls{ehvi} criterion (\ref{eq-ehvi-noise}). 
This implies iterating the expectation over the posterior $p\left(\boldsymbol{f}\left(X_{n}\right) \mid \mathcal{D}_{n}\right)$ 
of the function values at previously evaluated data points $X_{n}$, given the noisy observations $\mathcal{D}_{n}=\left\{\boldsymbol{x}_{i}, \boldsymbol{y}_{i},\left(\Sigma_{i}\right)\right\}_{i=1}^{n}$:

\begin{equation}
    \label{eq-ehvi-noise}
    \begin{aligned}
        & \alpha_{\mathrm{NEHVI}}(\boldsymbol{x})=\int \alpha_{\mathrm{EHVI}}\left(\boldsymbol{x} \mid \mathcal{P}_{n}\right) p\left(\boldsymbol{f} \mid \mathcal{D}_{n}\right) d \boldsymbol{f}; \\ 
        &\hat{\alpha}_{\mathrm{NEHVI}}(\boldsymbol{x})=\frac{1}{N} \sum_{i=1}^{N} \operatorname{HVI}\left(\tilde{\boldsymbol{f}}_{i}(\boldsymbol{x}) \mid \mathcal{P}_{i}\right),
    \end{aligned}
\end{equation}
where $P_{i}$ denotes the \gls{pf} over $\boldsymbol{f}\left(X_{i}\right)$, $\tilde{\boldsymbol{f}}_{i}\left(X_{n}, \boldsymbol{x}\right) \sim p\left(\boldsymbol{f}\left(X_{n}, \boldsymbol{x}\right) \mid \mathcal{D}_{n}\right)$ is a joint posterior and $\hat{\alpha}_{\text{NEHVI}}(\boldsymbol{x})$ is an approximation of $\alpha_{\text{NEHVI}}(\boldsymbol{x})$ by \gls{mc} integration.
Note that in practice, the algorithm is computationally complex.
Despite this, our interest lies primarily in the ability to support high-dimensional data \gls{pc}, which naturally falls into the domain of \gls{bo} with a noisy-\gls{ehvi} acquisition function.

\section{Simulation Results}\label{simres}

\subsection{Experimental Design}\label{simres-0}
To demonstrate the performance of the proposed \gls{mo} \gls{bo} methodology, we consider three cases of \gls{cf} networks, and one with a \gls{tdd} link between a single \gls{ue} and \gls{ap}.
In all settings, the \glspl{ap} are equipped with a 128 antenna \gls{mimo} system, while the \gls{ue}s operate with single antennas.
All units in the network are placed randomly with the following minimal distances between pairs: \gls{ap}-\gls{ue} -- 40 m, \gls{ue}-\gls{ue} -- 5 m, \gls{ap}-\gls{ap} -- 100 m.
Similarly to \cite{nguyenPilotAssignmentJoint2021}, 
there are $K$ orthogonal pilot signals and $P_{\max , l, k}^{\text{UL}} = 200$ mW with $P_{\max , l, k}^{\text{DL}} = 200K$ mW are set to ensure an equal total power budget for the \gls{ul} and \gls{dl} data transmissions.
The covariance matrix of channel $\mathbf{R}_{l, k}$  (\ref{eq-chan-mat}) is defined according to the exponential correlation model for a uniform linear array:

\begin{equation}
    \label{eq-chan-mat}
    \begin{aligned}
        & \mathbf{R}_{l, k}=\beta_{l, k}\left[\begin{array}{ccc}
            1 & \cdots & \left(r_{l, k}^{*}\right)^{M-1} \\
            \vdots & \ddots & \vdots \\
            \left(r_{l, k}\right)^{M-1} & \cdots & 1
            \end{array}\right]; \\
        & \beta_{l, k}[\mathrm{~dB}]=-148.1-37.6 \log _{10}\left(d_{l, k} / 1 \mathrm{~km}\right)+z_{l, k},
    \end{aligned}
\end{equation}

\noindent where $\mu e^{j \theta_{l, k}}$ denotes the spatial correlation with $\mu$ in a closed unit interval, $\beta_{l, k}^{j}$ [dB] denotes the large-scale fading coefficients, $d_{l, k}$ denotes distance between respective \gls{ue} and \gls{ap}, and lastly $z_{l, k}$ is the log-normally distributed shadow fading term.
Since we focus on the task of data \gls{pc}, the pilot assignment is done according to the similarity between covariance matrices \cite{youPilotReuseMassive2020}.

\subsection{Pareto Frontiers}\label{simres-1}
To demonstrate the performance of a parallel \gls{mo} \gls{bo} scheme with \gls{ehvi} \cite{daultonParallelBayesianOptimization2021} and gain insight into our multi-objective problem (\ref{eq-problem-1}) at scale, we first consider three controllable experiments, which consist of finding \glspl{pf} for a \gls{tdd} link consisting of one \gls{ue} and one \gls{ap}.

\begin{figure*}
    \centerline{
        \includegraphics[scale=0.23]{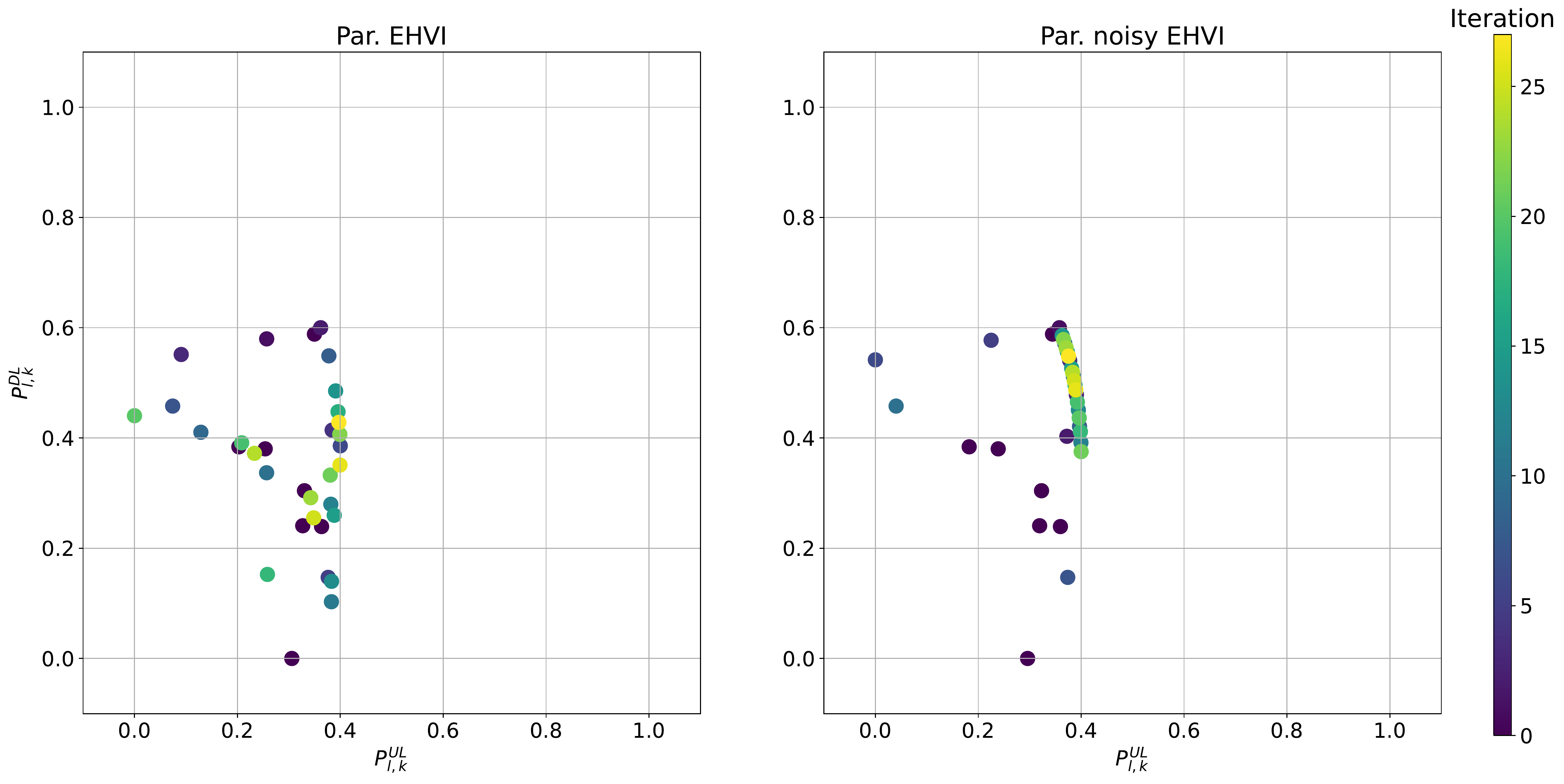}
        \includegraphics[scale=0.33]{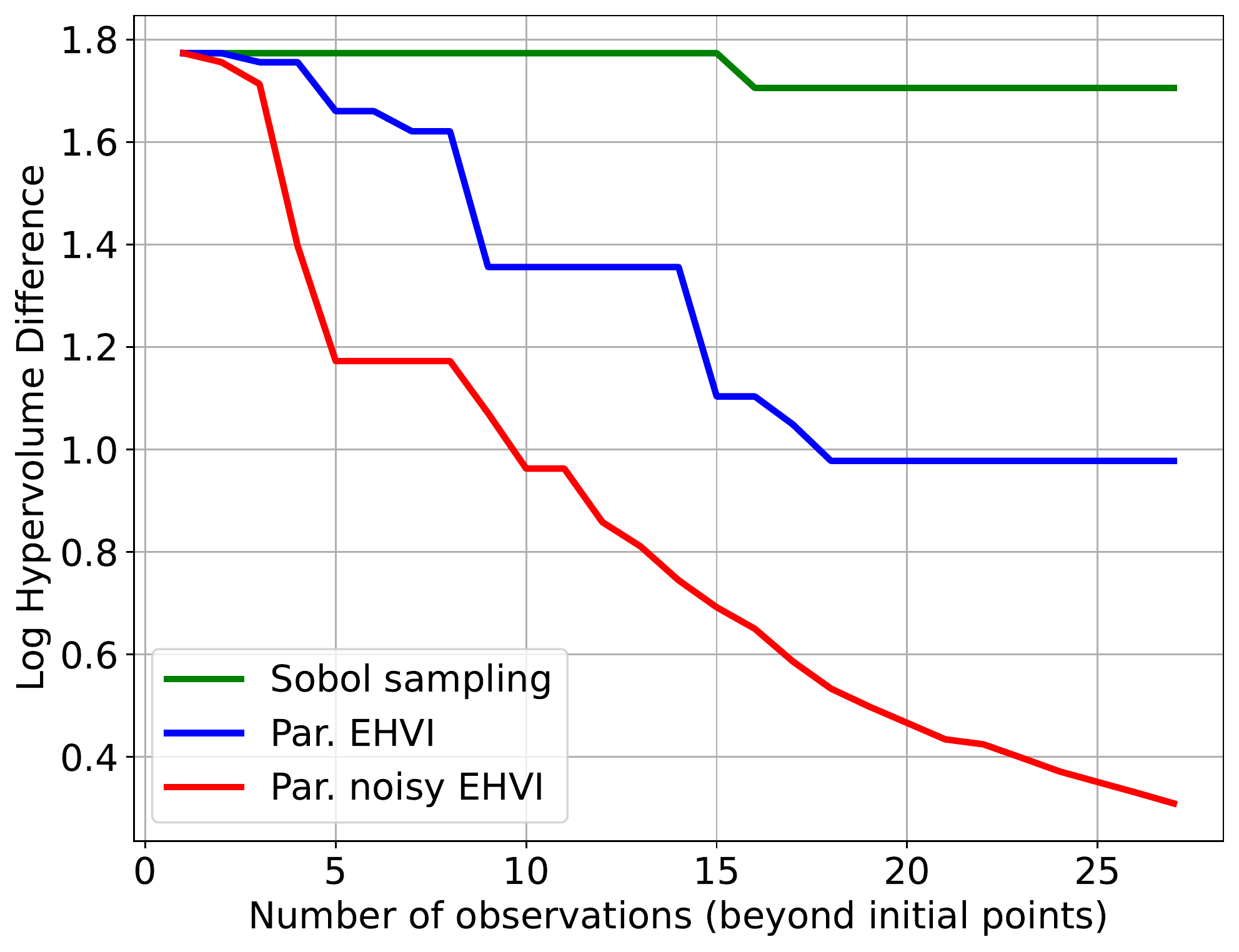}}
    \caption{\Glspl{pf} for $P_{l, k}^{\text{DL}}$ and $P_{l, k}^{\text{UL}}$ using \gls{ehvi} with log-hypervolume difference convergence plot, 1 \gls{ue} -- 1 \gls{ap}. }
    \label{fig-res-a}
\end{figure*}

\begin{figure*}
    \centerline{
        \includegraphics[scale=0.23]{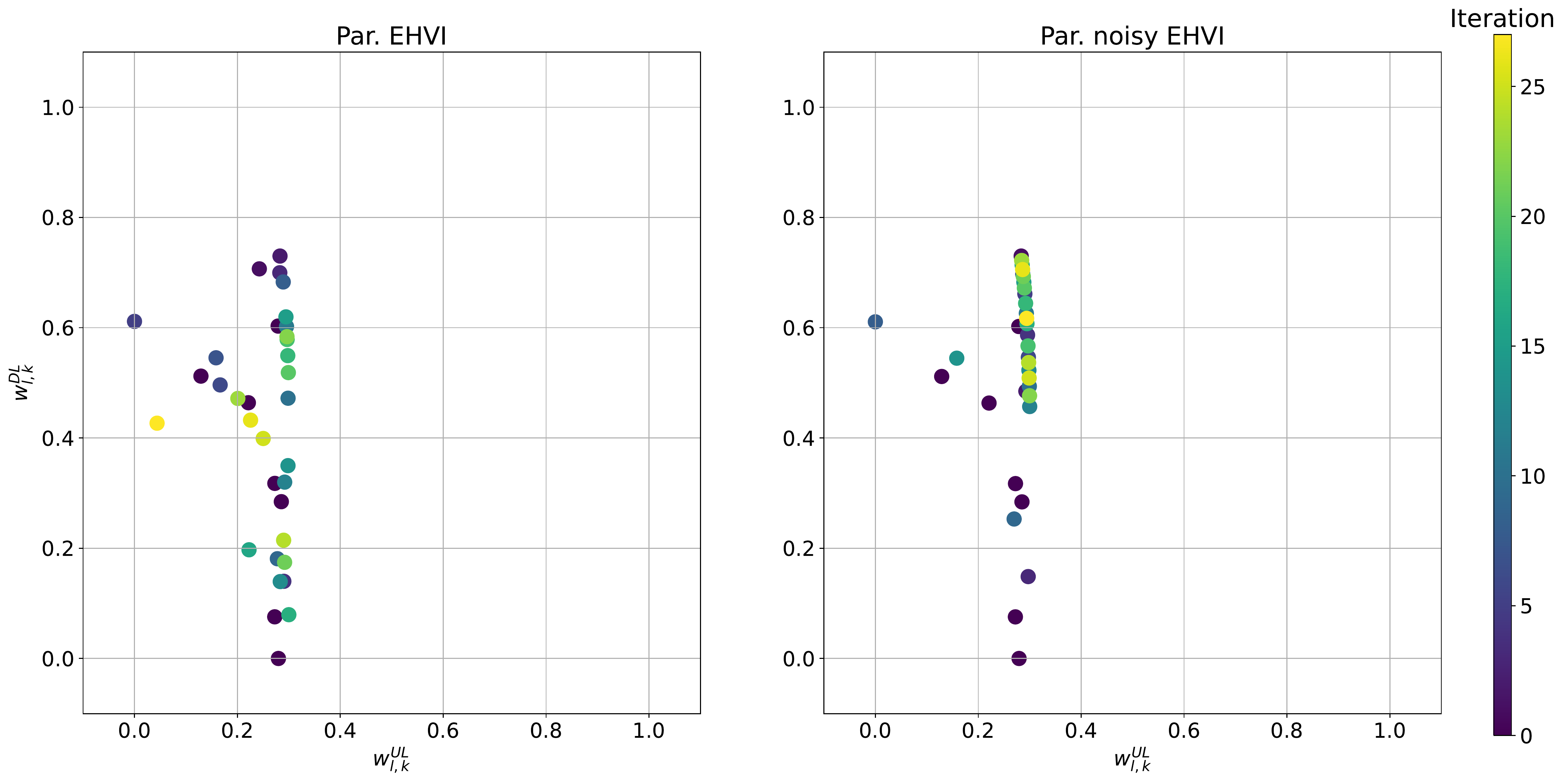}å
        \includegraphics[scale=0.33]{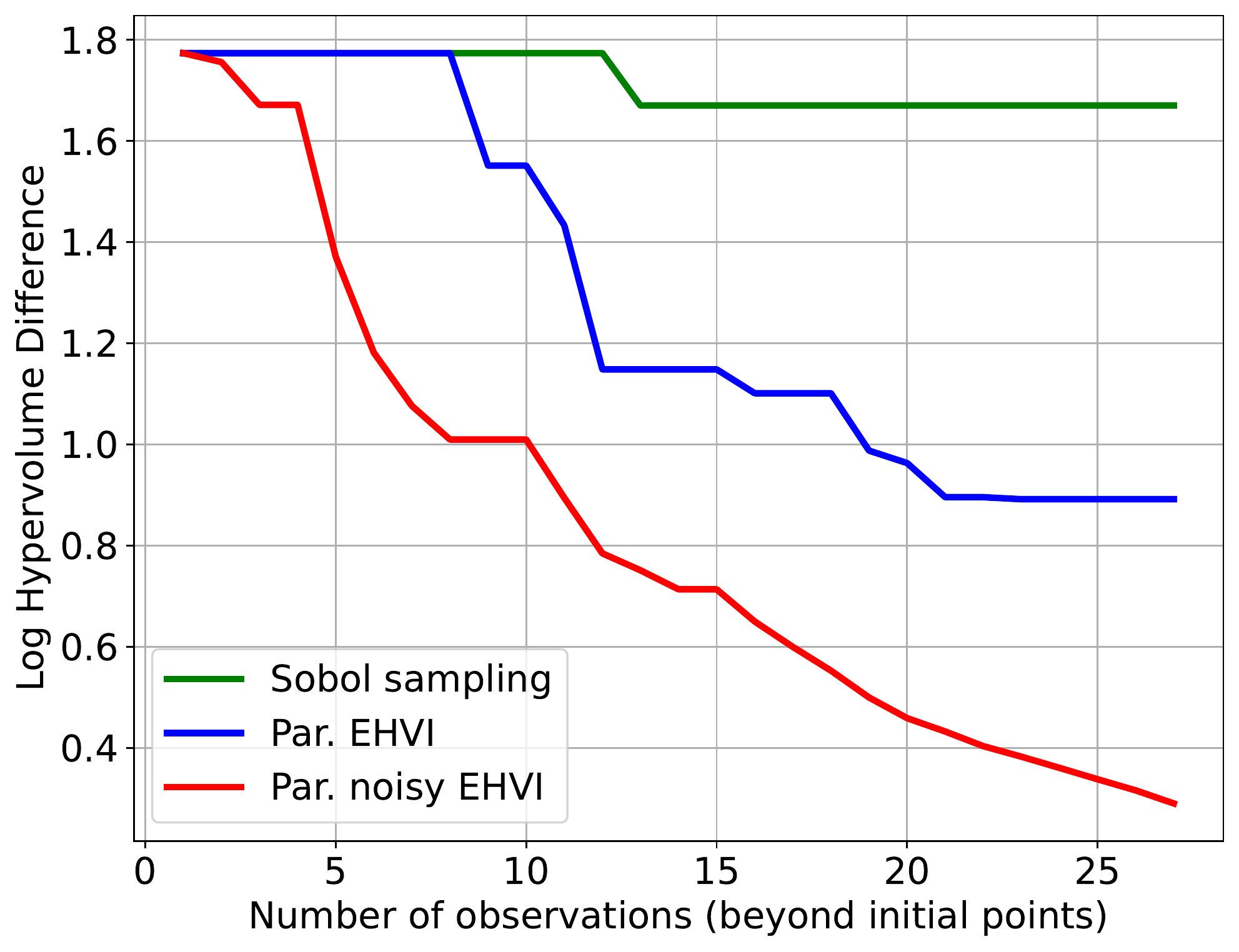}}
        \caption{\Glspl{pf} for $w_{l, k}^{\text{DL}}$ and $w_{l, k}^{\text{UL}}$ using \gls{ehvi} with log-hypervolume difference convergence plot, 1 \gls{ue} -- 1 \gls{ap}. }
        \label{fig-res-b}
\end{figure*}

\begin{figure*}
    \centerline{
        \includegraphics[scale=0.23]{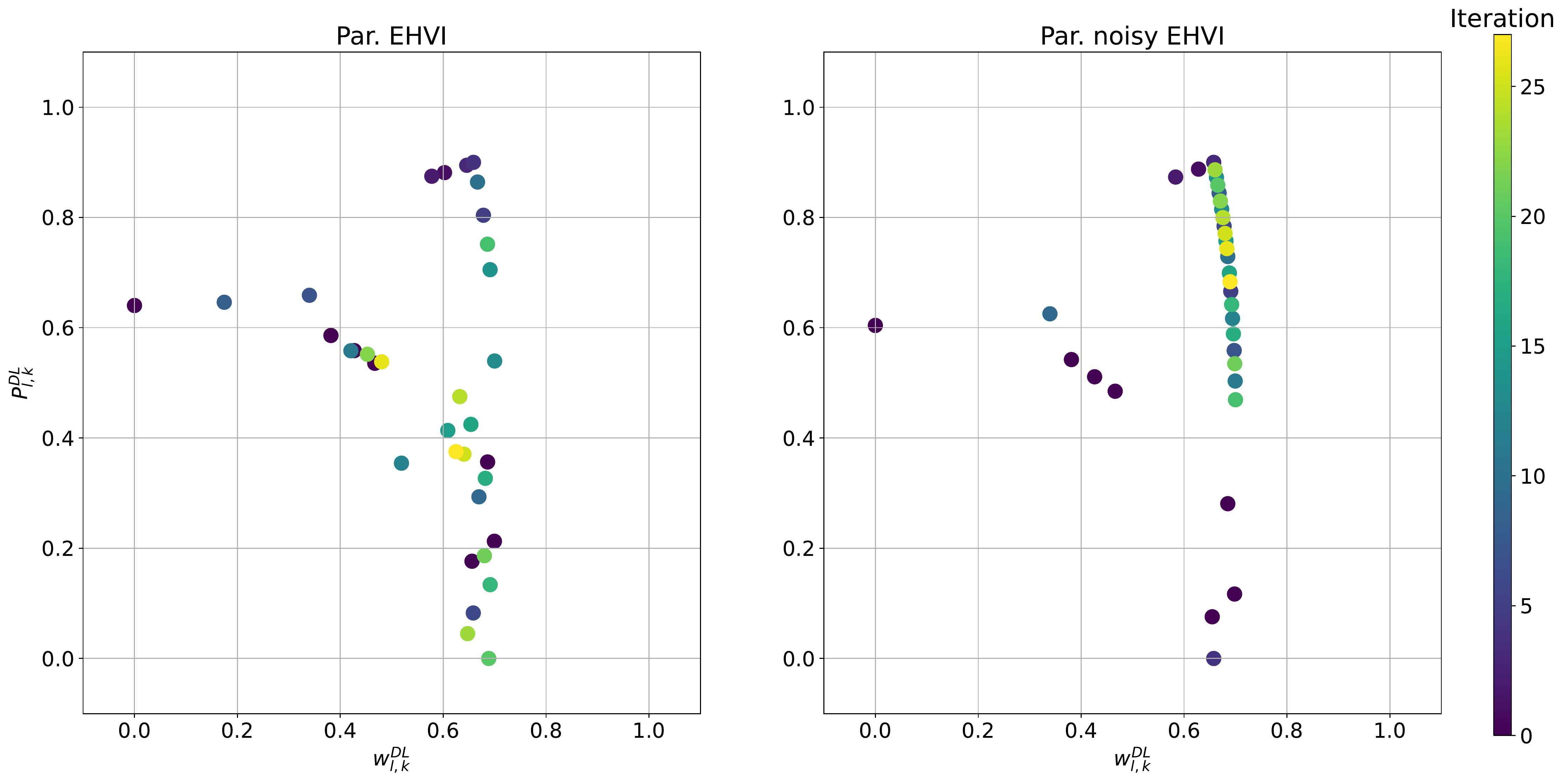}
        \includegraphics[scale=0.33]{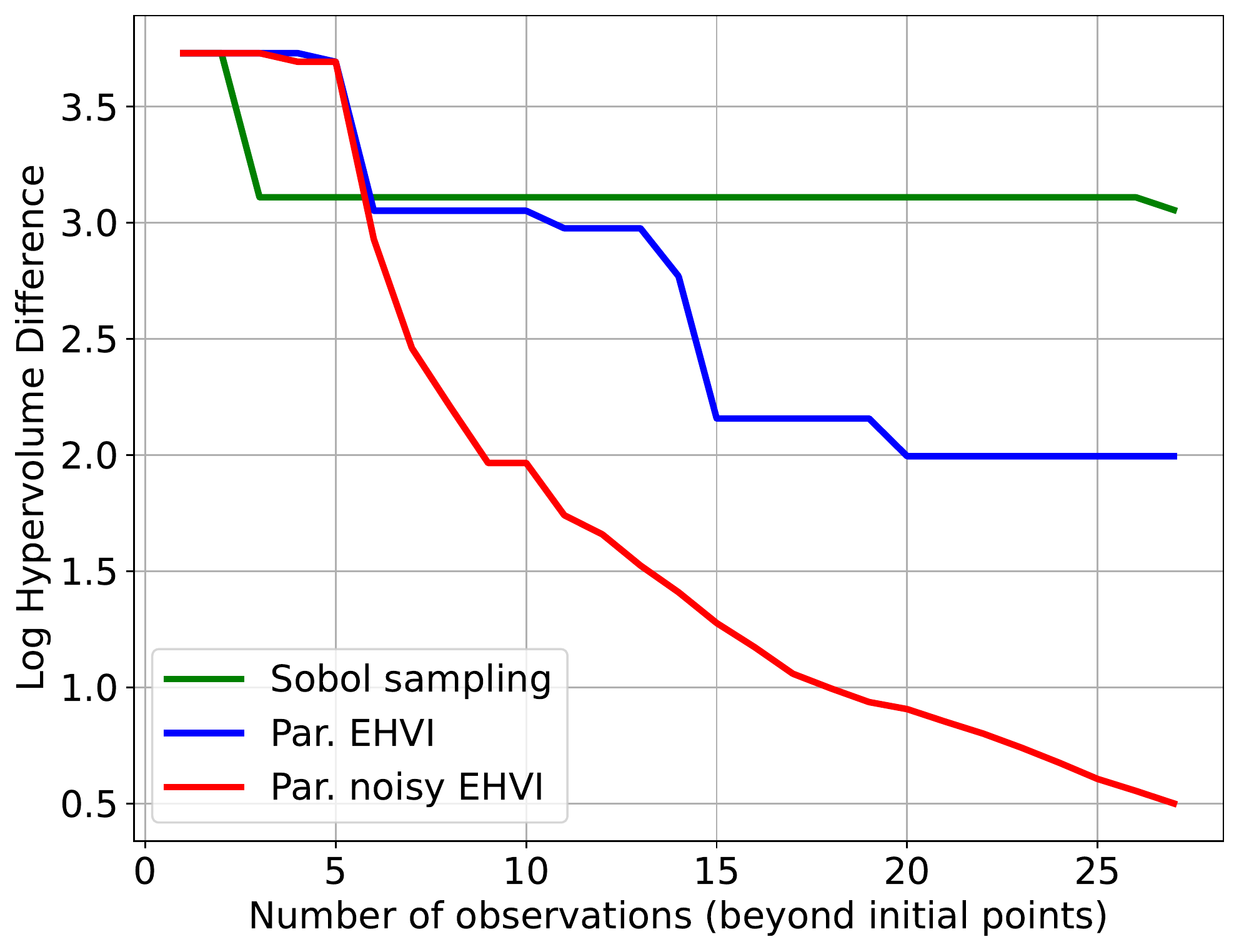}}
        \caption{\Glspl{pf} for $w_{l, k}^{\text{DL}}$ and $P_{l, k}^{\text{DL}}$ using \gls{ehvi} with log-hypervolume difference convergence plot, 1 \gls{ue} -- 1 \gls{ap}. }
    \label{fig-res-c}
\end{figure*}

To this end, we first fix $w_{1, 1}^{\mathrm{DL}}$ and $w_{1, 1}^{\mathrm{UL}}$ by assigning random values to them, and optimise $P_{1, 1}^{\mathrm{DL}}$ and $P_{1, 1}^{\mathrm{UL}}$.
Fig. (\ref{fig-res-a}) corresponds to this experiment, and contains the estimated \glspl{pf} for \gls{ehvi} and noisy-\gls{ehvi}, as well as the convergence curves for the log-hypervolume difference, where they are compared with quasi-\gls{mc} sampling, based on Sobol sequences.
Next, in Fig. (\ref{fig-res-b}) and Fig. (\ref{fig-res-c}) we reverse the problem by optimising the pairs: $w_{1, k}^{\mathrm{DL}}$ with $w_{1, 1}^{\mathrm{UL}}$ and $P_{1, k}^{\mathrm{DL}}$ with $w_{1, 1}^{\mathrm{DL}}$, respectively.

The \Glspl{pf} contain the collected observations for both \gls{ehvi} versions, where the colours correspond to the \gls{bo} iteration, at which the point was collected. 
Across all Fig. \ref{fig-res-a}-\ref{fig-res-c}, the central \glspl{pf} for noisy-\gls{ehvi} show that it is able to quickly identify the \gls{pf} and most of its evaluations are very close to it.
It is interesting to note that conventional \gls{ehvi} also identifies multiple observations close to the \gls{pf}, but due to relying on the technique of optimising random scalarisations, the shapes of \glspl{pf} are less defined.
Finally, notice that \gls{mc} sampling is unable to converge to a \gls{pf}, which is indicated by a stagnant log-hypervolume difference across all cases.
It is worth mentioning that the noisy-\gls{ehvi} continues to converge, and has a potential to approximate more complex \gls{pf}, which aligns with its convergence properties, as established in \cite{daultonParallelBayesianOptimization2021}.

Further controllable experiments (not reported here) provide additional intuition into performance of the chosen \gls{mo} optimisation methods, but our main goal is to maximise total \gls{se} by solving data \gls{pc}.


\subsection{Spectral Efficiencies}
In the key set of results reported below, we provide convergence plots with uncertainty estimates for \gls{se} maximisation via data \gls{pc} in three scenarios with increasing \gls{cf} network dimensionality.
Similarly to sub-section (\ref{simres-0}), we also compare quasi-\gls{mc} sampling and two \gls{ehvi} variants \cite{daultonParallelBayesianOptimization2021}.

\begin{figure*}[t]
    \begin{subfigure}[t]{0.33\textwidth}
      \includegraphics[scale=0.3]{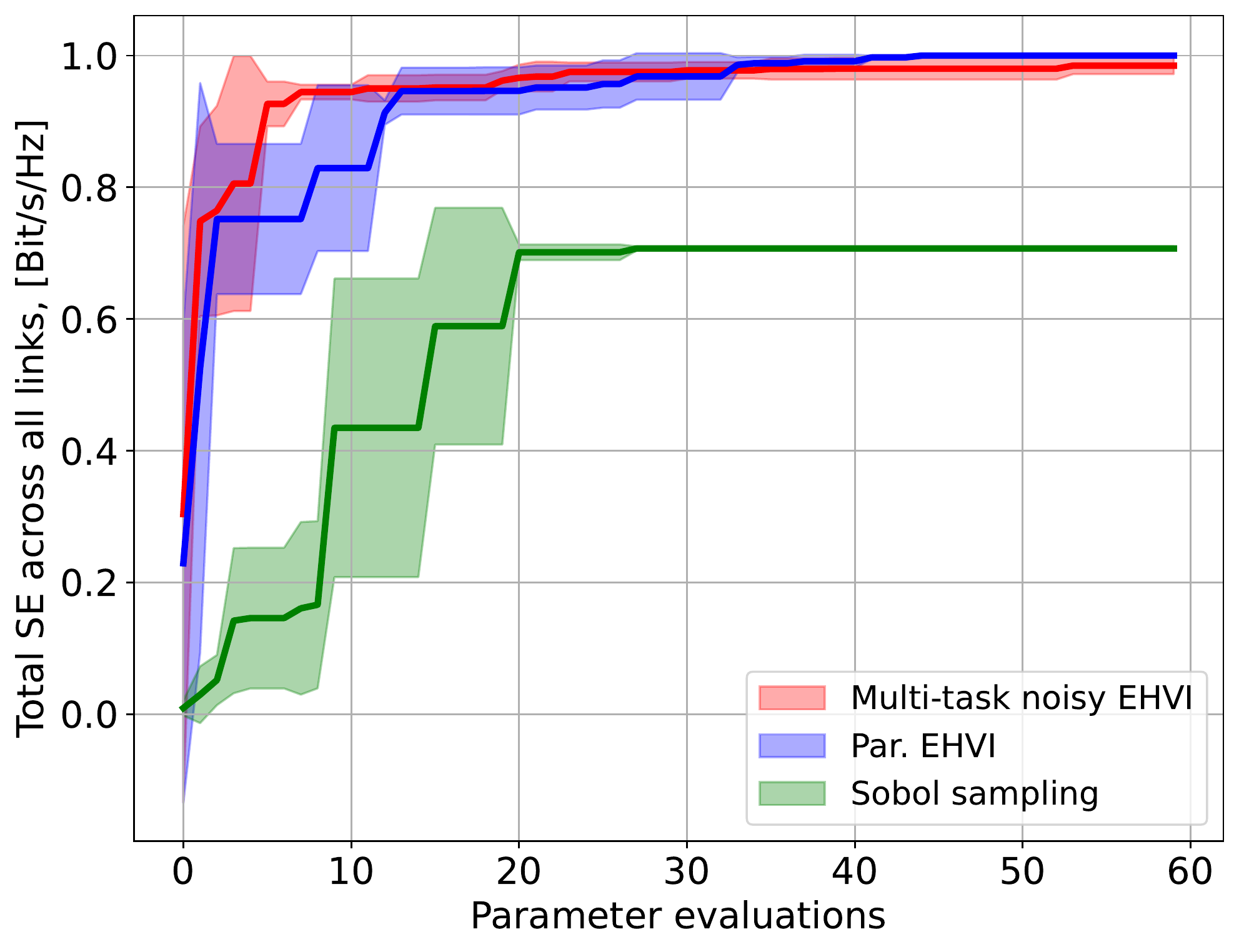}
      \caption{5 \gls{ue}, 5 \gls{ap}}
      \label{fig-res-se-a}
    \end{subfigure}
    \begin{subfigure}[t]{0.33\textwidth}
        \includegraphics[scale=0.3]{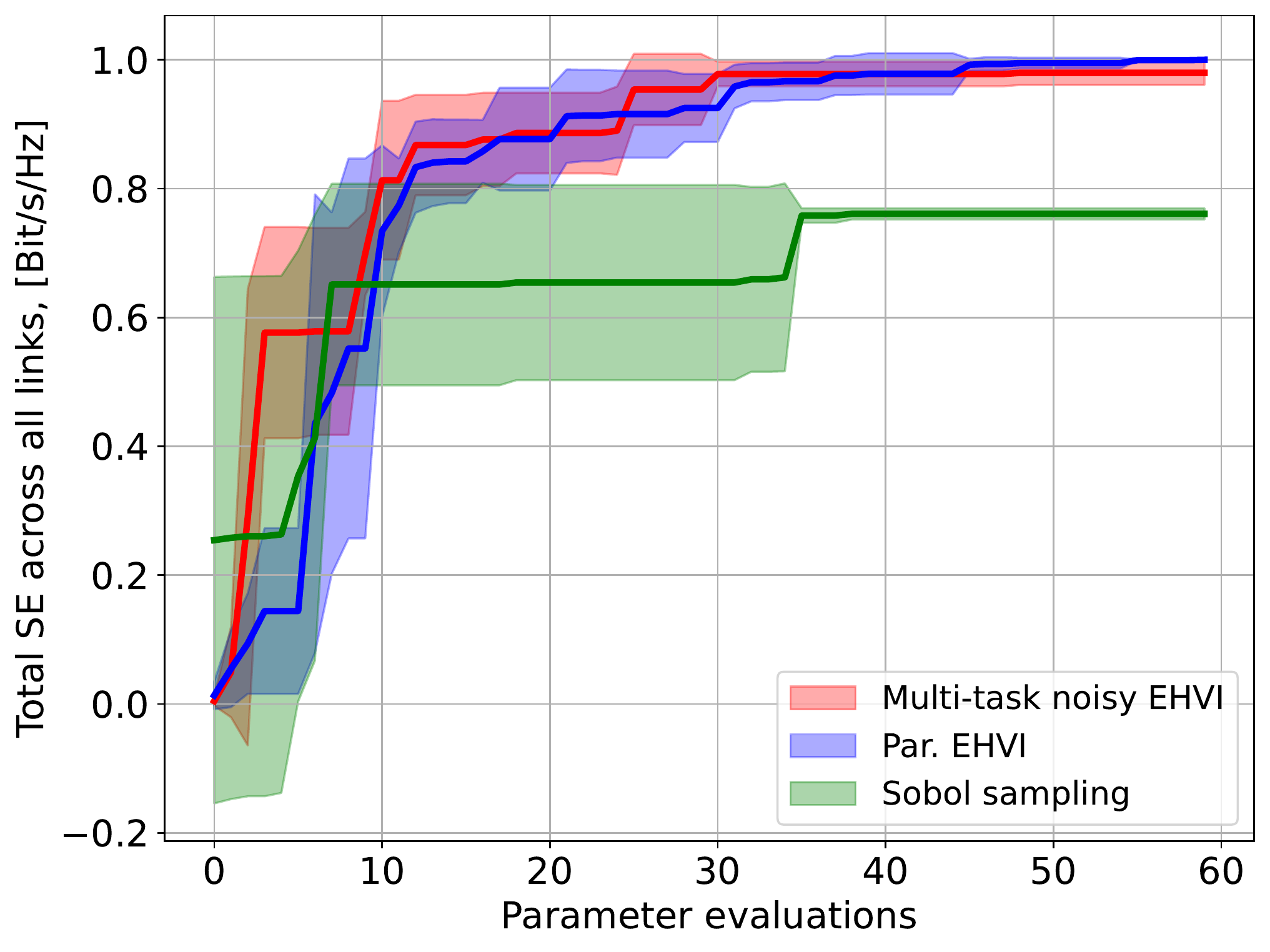}
        \caption{30 \gls{ue}, 20 \gls{ap}}
        \label{fig-res-se-b}
    \end{subfigure}
    \begin{subfigure}[t]{0.33\textwidth}
        \includegraphics[scale=0.3]{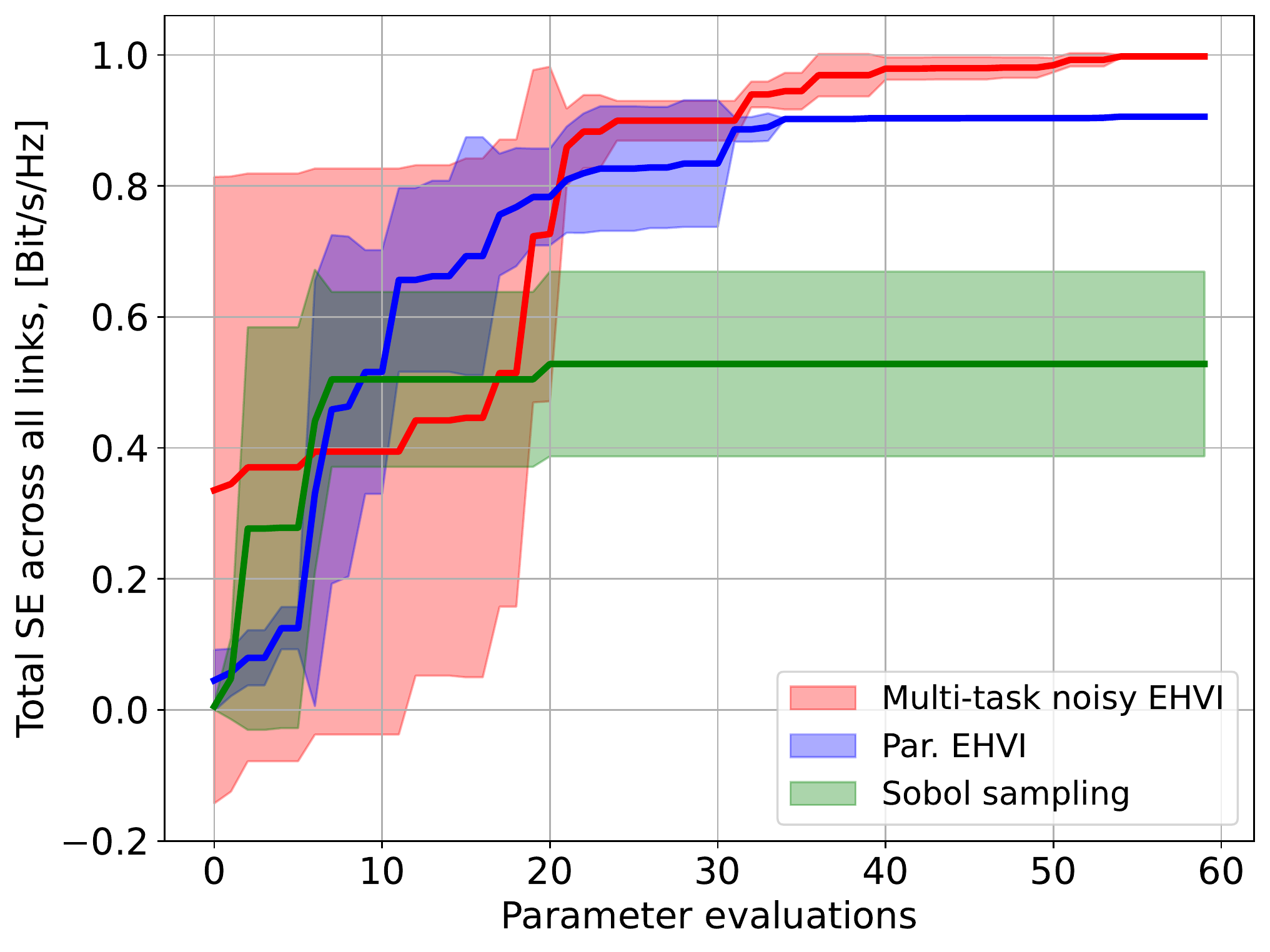}
        \caption{60 \gls{ue}, 40 \gls{ap}}
        \label{fig-res-se-c}
    \end{subfigure}
    \caption{Convergence rates for total \glsentrylong{se} as a function of $w_{l, k}^{\alpha}$ and $p_{l, k}^{\alpha}$ for 3 network cases.}
\end{figure*}

Fig. \ref{fig-res-se-a}, \ref{fig-res-se-b} and \ref{fig-res-se-c} show the total or summed \gls{se} for all established \gls{ue}-\gls{ap} links in the network for problem (\ref{eq-problem-1}).
The normalisation is done in order to compare the convergence processes in relation with the number of parameter evaluations across all three network cases.

The first case only contains 5 \gls{ue}s and 5 \gls{ap}s, which is comparable to the number of units in previous works \cite{vanchienJointPilotDesign2018, nguyenPilotAssignmentJoint2021}.
Here, both versions of \gls{ehvi} show similar performance by reaching a \gls{pf} solution after 20 parameter evaluations.
The real benefit of noisy-\gls{ehvi} becomes apparent in large scenarios containing in total 50 and 100 units in the network.
There, conventional methods do not scale well and random sampling struggles to continue convergence.  
Fig. \ref{fig-res-se-c} confirms this notion, as quasi-\gls{mc} maintains high uncertainty through all evaluations, while the noisy-\gls{ehvi} continuously reduces.

Because studied problem is multi-objective and combinatorial in its nature, it is difficult to know the exact optimum.
In context of future ressearch, it would be crucial to utilise high interpretability of \gls{bo} in combination with \glspl{pf} in order to assess properties of solution on per \gls{ue} basis.

\section{Concluding Remarks and Future Research}
\label{closure}
In this paper, we have described and shown a solution to the data \gls{pc} problem in a \gls{cf} setting, using an advanced \gls{bo} engine of the multi-objective noisy-\gls{ehvi}.

Despite the fact that nature of the problem does not allow to explicitly analyse the convergence towards the global optimum or a unique \gls{pf}, \gls{mo} \gls{bo} provides 
various benefits, which make its application advantageous in the domain of \gls{rrm}.
For example, \gls{bo} can be used to emulate the behaviour of large scale dynamical networks, effectively replacing existing link-level simulators.
Another advantage of the Bayesian approach is its modularity, which allows to combine \gls{gp} nonparametric and parametric models and build statistical emulators.  

A statistical emulator is an effective data-driven model that learns about a given simulation. Most notably, it learns with uncertainty.
In practice, the emulators may replace simulators, and they can invoke the simulator to make updates.
A statistical emulator is also a system that reconstructs the simulation with a statistical model. 
Together with reconstruction, a statistical emulator can be used to correlate with the real environment as well as calibrate the simulation to the environment.
This allows the emulator to characterise where the simulation can be relied on and consequentially to adjudicate between simulations. This is known as multi-fidelity emulation. 

Based on the aforementioned properties and results, we envision a future work in the context of \gls{mimo} \gls{cf} network emulators, which will not be hindered by scalability or restrictions in terms of the tunable parameters. 

% ––– Print acknowledgement information ––––––––––––––––––––––––––––––
\section*{Acknowledgement}
The work of S. Tambovskiy is funded by the Marie Skłodowska Curie action WINDMILL (grant No. 813999). G. Fodor was supported by the Swedish Foundation for Strategic Research Grant for Future Software Systems (FuSS), Grant No. FUS21-0004.

% ––– Print bibliography ––––––––––––––––––––––––––––––
\bibliography{PIMRC}

\begin{thebibliography}{10}
\providecommand{\url}[1]{#1}
\csname url@samestyle\endcsname
\providecommand{\newblock}{\relax}
\providecommand{\bibinfo}[2]{#2}
\providecommand{\BIBentrySTDinterwordspacing}{\spaceskip=0pt\relax}
\providecommand{\BIBentryALTinterwordstretchfactor}{4}
\providecommand{\BIBentryALTinterwordspacing}{\spaceskip=\fontdimen2\font plus
\BIBentryALTinterwordstretchfactor\fontdimen3\font minus
  \fontdimen4\font\relax}
\providecommand{\BIBforeignlanguage}[2]{{%
\expandafter\ifx\csname l@#1\endcsname\relax
\typeout{** WARNING: IEEEtran.bst: No hyphenation pattern has been}%
\typeout{** loaded for the language `#1'. Using the pattern for}%
\typeout{** the default language instead.}%
\else
\language=\csname l@#1\endcsname
\fi
#2}}
\providecommand{\BIBdecl}{\relax}
\BIBdecl

\bibitem{ranjbarCellFreeMMIMOSupport2022}
V.~Ranjbar, A.~Girycki, M.~A. Rahman, S.~Pollin, M.~Moonen, and E.~Vinogradov,
  ``Cell-{{Free mMIMO Support}} in the {{O-RAN Architecture}}: {{A PHY Layer
  Perspective}} for {{5G}} and {{Beyond Networks}},'' \emph{IEEE Communications
  Standards Magazine}, vol.~6, no.~1, pp. 28--34, Mar. 2022.

\bibitem{chenSurveyUsercentricCellfree2021}
S.~Chen, J.~Zhang, J.~Zhang, E.~Bj{\"o}rnson, and B.~Ai, ``A survey on
  user-centric cell-free massive {{MIMO}} systems,'' \emph{Digital
  Communications and Networks}, Dec. 2021.

\bibitem{bragaJointPilotData2022}
I.~M. Braga, R.~P. Antonioli, G.~Fodor, Y.~C.~B. Silva, and W.~C. Freitas,
  ``Joint {{Pilot}} and {{Data Power Control Optimization}} in the {{Uplink}}
  of {{User-Centric Cell-Free Systems}},'' \emph{IEEE Communications Letters},
  vol.~26, no.~2, pp. 399--403, Feb. 2022.

\bibitem{wuProportionalFairResourceAllocation2022}
S.~Wu, Y.~Wei, S.~Zhang, and W.~Meng, ``Proportional-{{Fair Resource
  Allocation}} for {{User-Centric Networks}},'' \emph{IEEE Transactions on
  Vehicular Technology}, vol.~71, no.~2, pp. 1549--1561, Feb. 2022.

\bibitem{nasirDeepReinforcementLearning2021}
Y.~S. Nasir and D.~Guo, ``Deep {{Reinforcement Learning}} for {{Joint
  Spectrum}} and {{Power Allocation}} in {{Cellular Networks}},'' in \emph{2021
  {{IEEE Globecom Workshops}} ({{GC Wkshps}})}, Dec. 2021, pp. 1--6.

\bibitem{saxenaReinforcementLearningEfficient2022}
V.~Saxena, H.~Tullberg, and J.~Jald{\'e}n, ``Reinforcement {{Learning}} for
  {{Efficient}} and {{Tuning-Free Link Adaptation}},'' \emph{IEEE Transactions
  on Wireless Communications}, vol.~21, no.~2, pp. 768--780, Feb. 2022.

\bibitem{girgisPredictiveControlCommunication2021}
A.~M. Girgis, J.~Park, M.~Bennis, and M.~Debbah, ``Predictive {{Control}} and
  {{Communication Co-Design}} via {{Two-Way Gaussian Process Regression}} and
  {{AoI-Aware Scheduling}},'' \emph{IEEE Transactions on Communications},
  vol.~69, no.~10, pp. 7077--7093, Oct. 2021.

\bibitem{vanchienJointPilotDesign2018}
T.~Van~Chien, E.~Bj{\"o}rnson, and E.~G. Larsson, ``Joint {{Pilot Design}} and
  {{Uplink Power Allocation}} in {{Multi-Cell Massive MIMO Systems}},''
  \emph{IEEE Transactions on Wireless Communications}, vol.~17, no.~3, pp.
  2000--2015, Mar. 2018.

\bibitem{nguyenPilotAssignmentJoint2021}
T.~H. Nguyen, T.~V. Chien, H.~Q. Ngo, X.~N. Tran, and E.~Bj{\"o}rnson, ``Pilot
  {{Assignment}} for {{Joint Uplink-Downlink Spectral Efficiency Enhancement}}
  in {{Massive MIMO Systems With Spatial Correlation}},'' \emph{IEEE
  Transactions on Vehicular Technology}, vol.~70, no.~8, pp. 8292--8297, Aug.
  2021.

\bibitem{wangInterSliceRadioResource2021}
T.~Wang and S.~Wang, ``Inter-{{Slice Radio Resource Allocation}}: {{An Online
  Convex Optimization Approach}},'' \emph{IEEE Wireless Communications},
  vol.~28, no.~5, pp. 171--177, Oct. 2021.

\bibitem{liDeepReinforcementLearningBased2021}
Y.~Li, J.~Jiang, C.~Jia, Y.~Yuan, Z.~Zhao, Y.~Du, and Z.~Wang, ``Deep
  {{Reinforcement Learning-Based Multi-Panel Beam Management}} in {{Massive
  MIMO Systems}}: {{Algorithm Design}} and {{System-Level Simulation}},'' in
  \emph{2021 {{IEEE}} 32nd {{Annual International Symposium}} on {{Personal}},
  {{Indoor}} and {{Mobile Radio Communications}} ({{PIMRC}})}, Sep. 2021, pp.
  1--6.

\bibitem{dreifuerstOptimizingCoverageCapacity2021}
R.~M. Dreifuerst, S.~Daulton, Y.~Qian, P.~Varkey, M.~Balandat, S.~Kasturia,
  A.~Tomar, A.~Yazdan, V.~Ponnampalam, and R.~W. Heath, ``Optimizing
  {{Coverage}} and {{Capacity}} in {{Cellular Networks}} using {{Machine
  Learning}},'' in \emph{{{ICASSP}} 2021 - 2021 {{IEEE International
  Conference}} on {{Acoustics}}, {{Speech}} and {{Signal Processing}}
  ({{ICASSP}})}, Jun. 2021, pp. 8138--8142.

\bibitem{maggiBayesianOptimizationRadio2021}
L.~Maggi, A.~Valcarce, and J.~Hoydis, ``Bayesian {{Optimization}} for {{Radio
  Resource Management}}: {{Open Loop Power Control}},'' \emph{IEEE Journal on
  Selected Areas in Communications}, vol.~39, no.~7, pp. 1858--1871, Jul. 2021.

\bibitem{ellerLocalizingBasestationsEndUser2022}
L.~Eller, V.~Raida, P.~Svoboda, and M.~Rupp, ``Localizing {{Basestations From
  End-User Timing Advance Measurements}},'' \emph{IEEE Access}, vol.~10, pp.
  5533--5544, 2022.

\bibitem{tekgulSampleEfficientLearningCellular2021}
E.~Tekgul, T.~Novlan, S.~Akoum, and J.~G. Andrews, ``Sample-{{Efficient
  Learning}} of {{Cellular Antenna Parameter Settings}},'' in \emph{2021 {{IEEE
  Information Theory Workshop}} ({{ITW}})}, Oct. 2021, pp. 1--6.

\bibitem{maddoxOptimizingHighDimensionalPhysics2021}
W.~Maddox, Q.~Feng, and M.~Balandat, ``Optimizing {{High-Dimensional Physics
  Simulations}} via {{Composite Bayesian Optimization}},''
  \emph{arXiv:2111.14911 [cs]}, Nov. 2021.

\bibitem{daultonParallelBayesianOptimization2021}
S.~Daulton, M.~Balandat, and E.~Bakshy, ``Parallel {{Bayesian Optimization}} of
  {{Multiple Noisy Objectives}} with {{Expected Hypervolume Improvement}},'' in
  \emph{Advances in {{Neural Information Processing Systems}}}, vol.~34.\hskip
  1em plus 0.5em minus 0.4em\relax {Curran Associates, Inc.}, 2021, pp.
  2187--2200.

\bibitem{wilsonMaximizingAcquisitionFunctions2018}
J.~Wilson, F.~Hutter, and M.~Deisenroth, ``Maximizing acquisition functions for
  {{Bayesian}} optimization,'' in \emph{Advances in {{Neural Information
  Processing Systems}}}, vol.~31.\hskip 1em plus 0.5em minus 0.4em\relax
  {Curran Associates, Inc.}, 2018.

\bibitem{gramacySurrogatesGaussianProcess2020}
R.~B. Gramacy, \emph{Surrogates: {{Gaussian Process Modeling}}, {{Design}}, and
  {{Optimization}} for the {{Applied Sciences}}}, 1st~ed.\hskip 1em plus 0.5em
  minus 0.4em\relax {Boca Raton}: {Chapman and Hall/CRC}, Jan. 2020.

\bibitem{youPilotReuseMassive2020}
L.~You and X.~Gao, ``Pilot {{Reuse}} for {{Massive MIMO}},'' in
  \emph{Encyclopedia of {{Wireless Networks}}}, X.~S. Shen, X.~Lin, and
  K.~Zhang, Eds.\hskip 1em plus 0.5em minus 0.4em\relax {Cham}: {Springer
  International Publishing}, 2020, pp. 1071--1073.

\end{thebibliography}

% ––– EOF ––––––––––––––––––––––––––––––
\end{document}